\documentclass[useAMS,usenatbib,usegraphicx]{mn2e}

\title[Photometry of open star clusters] {Photometric study of open clusters Berkeley 96,
Berkeley 97, King 12, NGC 7261, NGC 7296, and NGC 7788}
\author[E.V. Glushkova et al.]{E.V.~Glushkova,$^{1,2}$\thanks{E-mail:
elena@sai.msu.ru} M.V.~Zabolotskikh,$^2$ S.E.~Koposov,$^{2,3}$
O.I.~Spiridonova,$^4$
\newauthor
S.I.~Leonova,$^{2}$ V.V.~Vlasyuk $^4$, and A.S.~Rastorguev$^{1,2}$\\
$^1$Faculty of Physics, Lomonosov Moscow State University, Moscow, 119992
Russia\\ $^2$Sternberg Astronomical Institute, Lomonosov Moscow State University, Universitetsky pr.
13, Moscow, 119992 Russia\\ $^3$Institute of Astronomy, University of Cambridge,
Madingley Road, Cambridge, CB3 0HA UK\\
$^4$Special Astrophysical Observatory, Russian Academy of Sciences,
Nizhnii Arkhyz, Karachai-Cherkessian Republic, 357147 Russia}

\begin{document}

\date{Accepted 2012 December 00. Received 2012 December 00; in original form 2012 December 00}

\pagerange{\pageref{firstpage}--\pageref{lastpage}} \pubyear{2013}

\maketitle

\label{firstpage}

\begin{abstract}
We present $BVR_cI_c$ CCD photometry in the fields of six Galactic
open clusters toward the Perseus spiral arm. These data,
complemented with $J$, $H$, and $K_S$ magnitudes from 2MASS, have been used to determine the ages,
distances, and colour excesses $E(B-V)$ for these clusters: 40 Myr,
$3180^{+440}_{-380}$ pc, $0.54 \pm 0.03$ mag (Berkeley 96); 250
Myr, $2410^{+220}_{-200}$ pc, $0.77 \pm 0.06$ mag (Berkeley 97);
70 Myr, $2490^{+180}_{-170}$ pc, $0.51 \pm 0.05$ mag (King 12);
160 Myr, $2830^{+160}_{-150}$ pc, $0.88 \pm 0.09$ mag (NGC 7261);
280 Myr, $2450^{+190}_{-170}$ pc, $0.24 \pm 0.03$ mag (NGC 7296); and
160 Myr, $2750^{+220}_{-210}$ pc, $0.49 \pm 0.02$ mag (NGC 7788).
We found gaps in the mass function of clusters Be 97, King 12, and
NGC 7788 in the mass intervals of [1.3--1.5], [1.4--1.6], and
[1.5--1.7] solar masses, respectively.

\end{abstract}


\section{Introduction}

Despite important progress in the study of the structure and
evolution of the Galactic disk, many questions remain unanswered, such as the location of inner spiral arms or determination of
the corotation distance. Open star clusters (OCLs) play the most important
role in the investigation of Galactic disk properties, because
their ages, distances, and colour excesses can be easily evaluated
from photometric CCD observations. However, the number of investigated OCLs
with the heliocentric distances above 2 kpc is still small. We
developed and carry out a program of investigation of the Perseus
spiral arm based on the space and age distributions of open
clusters. We already obtained BVRI CCD photometry and estimated the
main physical parameters of six OCLs in this region: King 13, King
18, King 19, King 20, NGC 136, and NGC 7245
(\citet{Glushkova2010}).

This study is concerned with another sample of six open clusters located
toward the Perseus spiral arm: Berkeley 96 ($l = 103\fdg72$, $b =
-2\fdg09$), Berkeley 97 ($l = 106\fdg66$, $b = 0\fdg38$), King 12
($l = 116\fdg12$, $b = -0\fdg13$), NGC 7261 ($l = 104\fdg04$, $b =
0\fdg91$), NGC 7296 ($l = 101\fdg88$, $b = -4\fdg60$), and NGC 7788
($l = 116\fdg43$, $b = -0\fdg78$).


The WEBDA database on open star clusters
\footnote{http://www.univie.ac.at/webda} contains the following
information on photometry of the clusters under study.

{\bf Berkeley 96.} \citet{Rio1984} measured photoelectric $UBV$
magnitudes of 10 stars and derived the cluster distance $r = 5.3$
kpc and the colour excesses  $E(B-V) = 0.68$  and $E(U-B) = 0.50$ mag.

{\bf Berkeley 97.} \citet{Tadross2008} evaluated the distance to the cluster $r = 1.8$ kpc,
its age $t \approx 200$ Myr, and colour excess $E(B-V) = 0.75$ mag, using $JHK_s$
magnitudes from 2MASS Point Source Catalog \citep{2MASS} exclusively.

{\bf King 12.} \citet{MohanPandey1984} obtained photoelectric
$UBV$ magnitudes for 30 stars and found the cluster distance to be $r =
2.49$ kpc and its colour excess $E(B-V)$ to vary from 0.52 to 0.69
mag (differential reddening). \citet{Haug1970} measured
photoelectric $UBV$ magnitudes of four stars in the cluster field.

{\bf NGC 7261.} Based on $UBV$ photographic magnitudes of 147
stars, \citet{Fenkart1968} determined the cluster age $t \approx
10$ Myr, distance $r = 3230$ pc, and colour excesses $E(B-V) = 1.00$
and $E(U-B) = 0.73$ mag. Using $UBViyz$ photoelectric photometry
of five K-giants, \citet{JennensHelfer1975} found the cluster age
$t \approx 200$ Myr, distance $r = 2200$ pc, metal abundance
$[Fe/H] = -0.7$, and colour excess $E(B-V) = 0.48$ mag.
\citet{Hoag1961} obtained photoelectric $UBV$ photometry for 28
stars and photographic $UBV$ photometry for 53 stars;
\citet{Eggen1968} measured $UBV$ photoelectric magnitudes for two
stars; \citet{Kubinec1973} measured photographic BV magnitudes for
seven stars.

{\bf NGC 7296.} Based on $\Delta a$ and $BVR$ CCD photometry for
about 140 stars, \citet{Netopil2005} determined the cluster age $t
\approx 100$ Myr, distance $r = 2930$ pc, and colour excesses $E(B-V) =
0.15 \pm 0.02$ and $E(V-R) = 0.00 \pm 0.02$ mag.

{\bf NGC 7788.} \citet{Haug1970} measured photoelectric $UBV$
magnitudes of five stars; \citet{Becker1965} measured photographic
$UBV$ magnitudes of 113 stars; and \citet{Frolov1977} found
photographic $BV$ magnitudes of 67 stars.

The clusters Be 96, King 12, NGC 7261, and NGC 7788 were studied by means of photographic and photoelectric photometry down to V-band limiting magnitudes of 14.5--16.5 mag. According to the estimates reported by the above authors and by \citet{Dias2002} for NGC 7788, the distances to all these clusters exceed 2 kpc. Hence the published colour-magnitude diagram (CMD) isochrone fits involved only the upper parts of the main sequences of the clusters studied whose physical parameters could therefore be determined with large errors. Berkeley 97 was studied using  only $JHK_S$ data,  also down to a shallow limiting magnitude. The only  published color-magnitude diagram of NGC 7296 is based on CCD photometry and goes down to a V-band limiting magnitude of 16.8 mag. CCD  data is known to be  often fraught with systematic errors and we therefore decided to repeat the observations of this cluster. We also wanted to verify a very small value $E(V-R)=0.00$ derived by \citet{Netopil2005}. All these factors justify our choice of the six open clusters listed above.

\section{Observations and Data Reduction}

We observed six open clusters in the Johnson/Kron--Cousins broadband $BVR_cI_c$ 
using the Zeiss--1000 telescope at the Special Astrophysical
Observatory of the Russian Academy of Sciences equipped with a photometer
based on an EEV 42--40 2K $\times$ 2K CCD. The pixel size, readout noise, and gain were equal to 13.5
$\mu m$, $4e^{-}$, and $2.08e^{-}$ per ADU, respectively. The scale and field of view were equal to
$0\farcs207$/pixel and about $7
\arcmin$, and
$0\farcs238$/pixel and about $8 \arcmin$ during the 2003 and 2009 observations, respectively.
The scale and field of view size differences were due to changed photometer layout. Table 1 shows the log of
observations. Its columns contain dates of observation, object
names, air masses $X = \sec z$, and exposure times in each band.
Figure 1 shows the images of all six clusters under study taken in $V$--band with the
exposure 300 s with the limiting magnitude of about 21 mag.

\begin{table}
\caption{Log of observations}
\begin{tabular}{ccccc} \hline
Date           &Object     &$X$ &Exposure time, s\\
               &           &    &$B$, $V$, $R_c$, $I_c$\\
\hline
Aug. 27, 2003  &NGC 7788   &1.05&$300$, $300$, $300$, $300$\\
               &           &    &$60$,  $30$,  $30$,  $30$ \\
               &NGC 7790   &1.07&$300$, $100$, $100$, $100$\\
Aug. 29, 2003  &NGC 7296   &1.01&$300$, $300$, $300$, $300$\\
               &           &    &$300$, $300$, $300$, $300$\\
               &           &    &$30$,  $30$,  $15$,  $15$ \\
               &NGC 7790   &1.06&$300$, $100$, $100$, $100$\\
Oct. 15, 2009  &NGC 7790   &1.06&$300$, $150$, $150$, $150$\\
               &King 12    &1.05&$300$, $300$, $300$, $300$\\
               &           &    &$30$,  $30$,  $30$,  $30$ \\
               &NGC 7788   &1.05&$300$, $300$, $300$, $300$\\
               &           &    &$--$,  $30$,  $30$,  $30$ \\
               &NGC 7790   &1.06&$300$, $150$, $150$, $150$\\
               &NGC 7790   &1.09&$300$, $150$, $150$, $150$\\
Oct. 16, 2009  &NGC 6940   &1.05&$300$, $150$, $150$, $150$\\
               &Berkeley 96&1.05&$300$, $300$, $300$, $300$\\
               &           &    &$30$,  $30$,  $30$,  $30$ \\
               &Berkeley 97&1.04&$300$, $300$, $300$, $300$\\
               &           &    &$30$,  $30$,  $30$,  $30$ \\
               &NGC 7790   &1.05&$300$, $150$, $150$, $150$\\
               &NGC 7296   &1.06&$300$, $300$, $300$, $300$\\
               &           &    &$30$,  $30$,  $30$,  $30$ \\
               &NGC 7790   &1.05&$300$, $150$, $150$, $150$\\
               &NGC 7790   &1.09&$300$, $150$, $150$, $150$\\
               &NGC 7790   &1.10&$300$, $150$, $150$, $150$\\
Oct. 18, 2009  &NGC 7790   &1.06&$300$, $150$, $150$, $150$\\
               &Berkeley 97&1.06&$300$, $300$, $300$, $300$\\
               &NGC 7790   &1.05&$300$, $150$, $150$, $150$\\
               &King 12    &1.07&$300$, $300$, $300$, $300$\\
               &NGC 7790   &1.08&$300$, $150$, $150$, $150$\\
Oct. 19, 2009  &NGC 7790   &1.05&$300$, $150$, $150$, $150$\\
               &NGC 7790   &1.10&$300$, $150$, $150$, $150$\\
Nov. 18, 2009  &NGC 7790   &1.05&$300$, $150$, $150$, $150$\\
               &Berkeley 97&1.06&$400$, $400$, $--$,  $--$  \\
               &Berkeley 97&1.06&$400$, $400$, $--$,  $--$  \\
               &Berkeley 96&1.09&$300$, $300$, $300$, $--$\\
               &           &1.12&$30$,  $30$,  $30$,  $30$ \\
               &NGC 7790   &1.07&$300$, $150$, $150$, $150$\\
\hline
\end{tabular}
\end{table}

\begin{figure*}
\includegraphics{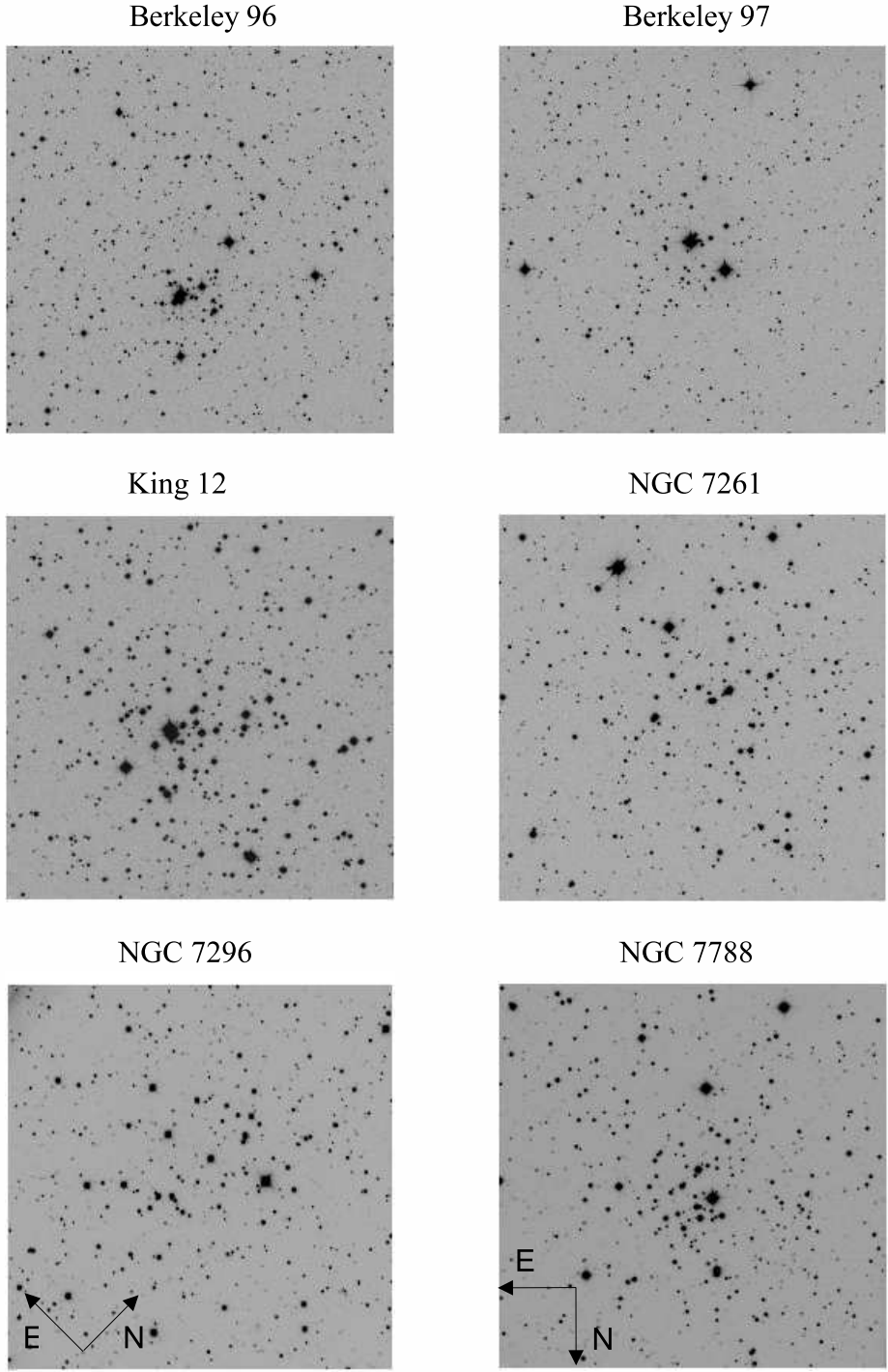} \caption{ $V$--band images of the
clusters studied (exposure time 300 s). The common orientation of
all frames (except for NGC 7296) relative to the equatorial
coordinate system is indicated by the arrows on the image of NGC
7788. The image size is near $8 \arcmin \times$ $8 \arcmin$ ($7 \arcmin \times$ $7 \arcmin$ for NGC
7296).} \label{images}
\end{figure*}

Primary data processing, image extraction, and analysis were
performed in the standard way using the ESO-MIDAS environment and
the DAOPHOT/ALLSTAR program package \citep{Stetson1987}. Prior to
the data reduction, we transformed the pixel coordinates for every
star into celestial coordinates for the epoch J2000. In addition
to the program clusters studied, the open cluster NGC 7790 was
observed several times for each night as a photometric standard,
and the NGC 6940 cluster was used once as a photometric standard (see Table 1). The standard magnitudes for stars in the
fields of NGC 7790 and NGC 6940 were taken from Stetson's database
\footnote{
http://http://www3.cadc-ccda.hia-iha.nrc-cnrc.gc.ca/community/STETSON/standards/}.
The ranges of magnitudes and colours for about 250 standard stars
in the cluster field are $12.7 < V < 19.1$ and $0.4 < (B-V) <
1.9$ mag, respectively.

We used the following formulae \citep{Hardie1964} to
transform the magnitudes and colours from the instrumental
to the standard system:
$$
V = v + \varepsilon(B-V) + \xi_{V_1},
$$
$$
V = v + \rho(V-R_c) + \xi_{V_2},
$$
$$
(B-V) = \mu(b-v) + \xi_{(B-V)},
$$
$$
(V-R_c) = \psi(v-r) + \xi_{(V-R_c)},
$$
$$
(V-I_c) = \varphi(v-i) + \xi_{(V-I_c)},
$$
where $b$, $v$, $r$, and $i$ are the magnitudes in the instrumental
system and $B$, $V$, $R_c$, and $I_c$ are the magnitudes in the
standard system. The formula for the $V$ magnitude as a function of
$(V-R_c)$ colour was used only for the stars without $B$ magnitudes. 
The averages of the median coefficients were $\varepsilon = -0.073
\pm 0.005$, $\rho = -0.125 \pm 0.012$, $\mu = 1.211 \pm 0.010$,
$\psi = 0.803 \pm 0.010$, and $\varphi = 0.875 \pm 0.007$ for
the observations done in 2003 and $\varepsilon = -0.076 \pm 0.003$,
$\rho = -0.131 \pm 0.005$, $\mu = 1.208 \pm 0.006$, $\psi = 0.910
\pm 0.006$, and $\varphi = 0.877 \pm 0.003$ for the observations done in
2009 (the $rms$ deviations are given as the errors). We used mean
values to recalculate the zero points $\xi_{V_1}$, $\xi_{V_2}$,
$\xi_{(B-V)}$,  $\xi_{(V-R_c)}$, and $\xi_{(V-I_c)}$ for the times
of observations of the standard clusters. The differences in the $\psi$ coefficients are presumably due to the changes in the photometer layout mentioned above.

As shown in Table 1, the standard clusters NGC 7790 and NGC 6940,
and all program clusters were observed at the same air mass $X$.
This explains why we did not estimate the
extinction coefficients. In the cases where NGC 7790 was observed
twice, before and after the program cluster, we calculated the zero
points of the transformation expressions by a linear interpolation
between zero point values calculated from the standard cluster.
Weighted mean $V$ magnitudes and $(B-V)$, $(V-R_c)$, and $(V-I_c)$
colours were computed and published only for those stars in the
fields of the six program clusters, for which the magnitudes and colours
derived from different frames and reduced to the standard system
differ by less than 0.1 mag. This value (0.1 mag) seems to be the typical magnitude error for the faintest measured stars and, at the same time, close to 3$\sigma$-variance of magnitudes measured on different frames. The final values of the errors in the magnitudes and colours were calculated as the
$rms$ errors of the weighted mean on all frames (taking into account the errors in the
aperture corrections and the errors introduced by the transformation
from the instrumental system to the standard one). However, if the
$rms$ deviation of the magnitude or colour calculated on all frames exceeded the $rms$ error of the weighted mean, this $rms$ deviation was taken as the final error.

The results of the photometric study of the clusters Berkeley 96,
Berkeley 97, King 12, NGC 7261, NGC 7296, and NGC 7788 are
presented in Tables 4, 5, 6, 7, 8, and 9,
respectively.
Tables 4--9 are available in the electronic form
only and are accessible at the CDS website \footnote{http://cdsarc.u-strasbg.fr/} and at the
Sternberg Astronomical Institute website \footnote{http://ocl.sai.msu.ru/}.
In this paper, we list a sample of few lines of Table 4 only as an example. Tables 5--9 have the same layout.
The tables include the equatorial coordinates $\alpha$ and
$\delta$ in degrees at the epoch J2000, $V$ magnitudes and
$(B-V)$, $(V-R_c)$, and $(V-I_c)$ colours and their errors $\sigma_V$,
$\sigma_{(B-V)}$, $\sigma_{(V-R_c)}$, and $\sigma_{(V-I_c)}$ for individual stars. We used 99.999 in all cells where the corresponding values are not available.

\section{Results and Discussion}
\subsection{Determination of Physical Parameters}

\begin{figure*}
\resizebox{\hsize}{!}{\includegraphics{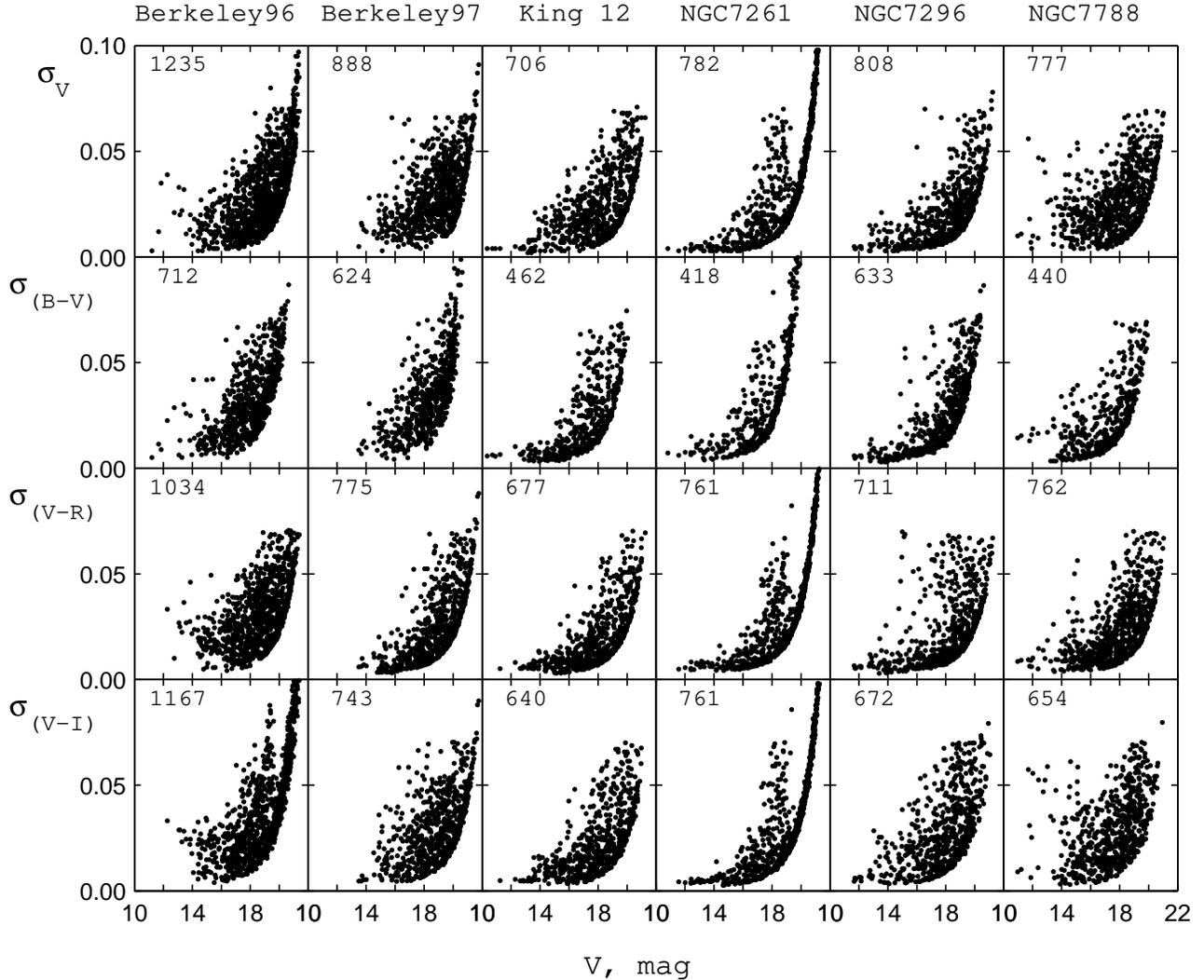}} \caption{ Errors
in the magnitudes, $\sigma_{V}$ , and in the colours, $\sigma_{(B-V)}$,
$\sigma_{(V-R_c)}$, and $\sigma_{(V-I_c)}$, versus $V$ magnitude for
the program clusters (from left to right) Berkeley 96, Berkeley 97,
King 12, NGC 7261, NGC 7296, and NGC 7788; the numbers on top of
each plot indicate the number of stars; all values along the
horizontal and vertical axes are given in magnitudes.}
\label{errors}
\end{figure*}

We derived $V$ magnitudes and $(B-V)$, $(V-R_c)$, and $(V-I_c)$ colours
for the stars in the fields of Berkeley 96, Berkeley 97, King 12,
NGC 7261, NGC 7296, and NGC 7788. In Fig.~\ref{errors}, the error
in the magnitude, $\sigma_{V}$ , and the errors in the colours,
$\sigma_{(B-V)}$, $\sigma_{(V-R_c)}$, and $\sigma_{(V-I_c)}$, are
plotted against the $V$ magnitude for each cluster; the numbers on
each plot indicate the number of stars. In most cases, the errors
do not exceed 0.10 mag; the magnitude limit is 20-21.5 mag,
depending on the band. The colour plots for NGC 7261 show narrow
``tails'', because only one frame in each band was taken with the
exposure 300 s, and no averaging was performed; the same is seen
for $\sigma_{(V-I_c)}$ (Berkeley 96).

\begin{figure*}
\resizebox{\hsize}{!}{\includegraphics{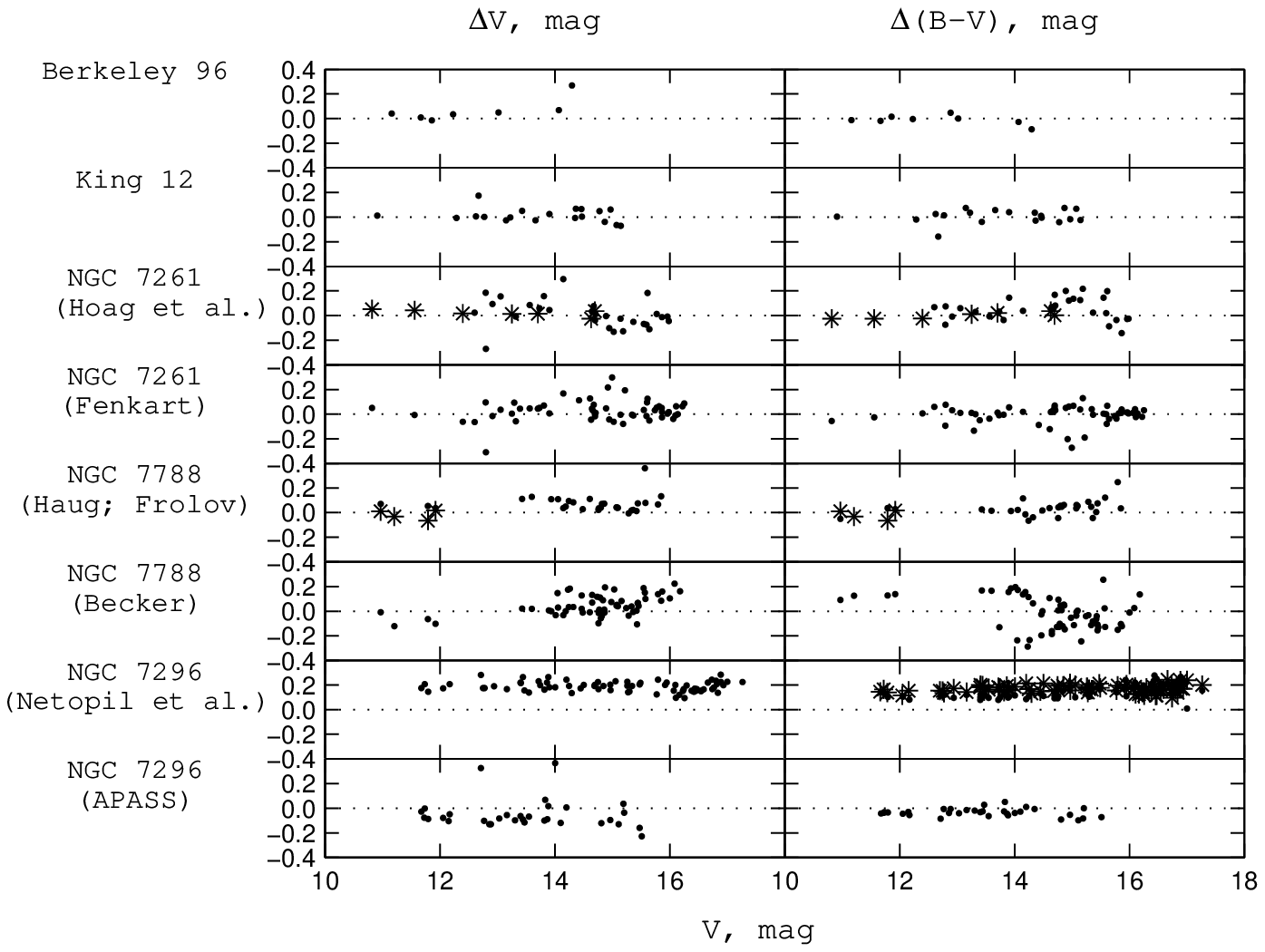}} \caption{
Differences of magnitudes, $\Delta V$ (left), and colours, $\Delta
(B-V)$ (right), for common stars in this paper and \citet{Rio1984}
in the field of Berkeley 96; in this paper and
\citet{MohanPandey1984} in the field of King 12; in this paper and
\citet{Hoag1961} (symbol ``star'' -- photoelectric photometry,
symbol ``dot'' -- photographic photometry), \citet{Fenkart1968} in
the field of NGC 7261; in this paper and \citet{Haug1970} (symbol
``star''), \citet{Frolov1977} (symbol ``dot''), \citet{Becker1965}
in the field of NGC 7788; in this paper and \citet{Netopil2005}
(symbol ``star'' -- $\Delta (B-V)$, symbol ``dot'' -- $\Delta
(V-R)$) in the field of NGC 7296; in this paper and the APASS catalogue in the field of NGC 7296. The V magnitude is along the
horizontal axis on all plots; the dashed line corresponds to zero
magnitude and colour differences.} \label{liter}
\end{figure*}

We compared the derived magnitudes and colours with published
photometry and show the magnitude and colour differences in
Fig.~\ref{liter}. Our photometric data are in good agreement with
all other photoelectric data - see plots for Be 96 and King 12,
``star'' symbols on the plots for NGC 7261 and NGC 7788. In
contrast, the comparison of our CCD photometry with photographic
values reported elsewhere exhibits rather wide spread, as shown by the ``dot''
symbols on the plots for NGC 7261 and NGC 7788. We also found 
systematic differences of approximately $0.2 \pm 0.05$ mag in $V$
and $0.1 \pm 0.05$ mag in $(V-R_c)$ between our values and those
derived by \citet{Netopil2005} for NGC 7296. Differences in the
$(B-V)$ colours show the systematic trend from $0.1$ to $0.25$ mag. However, 32 common stars identified in the APASS catalogue show small negative differences of about $-0.07 \pm 0.03$ mag in $V$ and $-0.04 \pm 0.02$ mag in $(B-V)$, respectively.

To estimate the cluster distance, age, and colour excess toward the
cluster, we built the following colour--magnitude diagrams (CMD) for
each program cluster: $(V,B-V)$, $(V,V-R_c)$, and $(V,V-I_c)$ -- from
our data, and $(J,J-H)$, $(K_s,J-K_s)$ -- from the 2MASS catalogue. The
faint stars with colour errors exceeding $0.05$ mag were excluded
from our consideration. We then fit each CMD to a solar-metallicity isochrone 
by \citet{Girardi2002} using the method
developed by \citet{Koposov2008}. In this method we superimposed the
isochrone on the colour--magnitude diagram in such a way that the
corresponding plot for the radial density distribution exhibits a
peak for the stars lying in the vicinity of the isochrone (the distance
in the colour index for such stars is less than $0.05$ mag) and an almost
flat distribution for all other stars (supposed field stars). The
centers of clusters were found from density maps built using data
from the 2MASS PSC. We considered the position of the
maximum of the density peak as the center of the cluster, and the
distance from the cluster center, at which the star density in the cluster region becomes
flat on the radial density distribution and equal to the density of field stars, as the cluster radius.
In this way we evaluated the distance, colour excess, and age for each
cluster on each CMD. To convert the colour excesses found from
different colour--magnitude diagrams to the colour excess $E(B-V)$
and to calculate the distance modulus, we used the following
relations: $A_{K_s} = 0.670 \cdot E(J-K_s)$ (\citet{Dutra2002}) and
$A_V = 3.08 \cdot E(B-V)$, $E(V-R_c)=0.61 \cdot E(B-V)$, $E(V-I_c)
= 1.35 \cdot E(B-V)$, $E(V-J) = 2.25 \cdot E(B-V)$, and $E(V-H) = 2.57
\cdot E(B-V)$ (\citet{He1995}).

\begin{figure*}
\resizebox{\hsize}{!}{\includegraphics{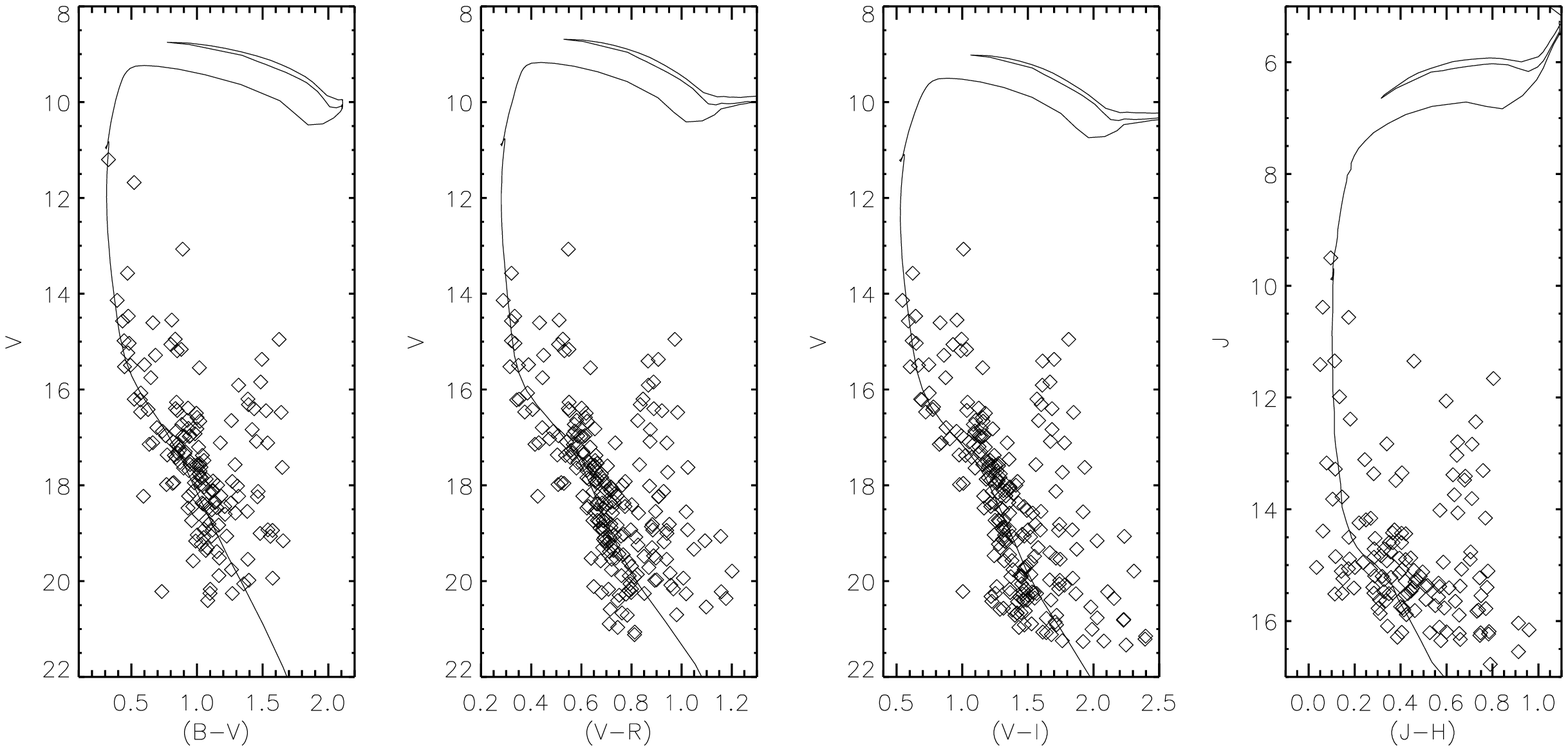}} \caption{
Colour--magnitude diagrams for Berkeley 96. The diamonds mark the
positions of stars on the diagram; the solid lines indicate the
corresponding isochrones. All values along the horizontal and
vertical axes are given in magnitudes. The size of the region for
which the diagram was constructed is $3 \arcmin$.} \label{izo1}
\end{figure*}

\begin{figure*}
\resizebox{\hsize}{!}{\includegraphics{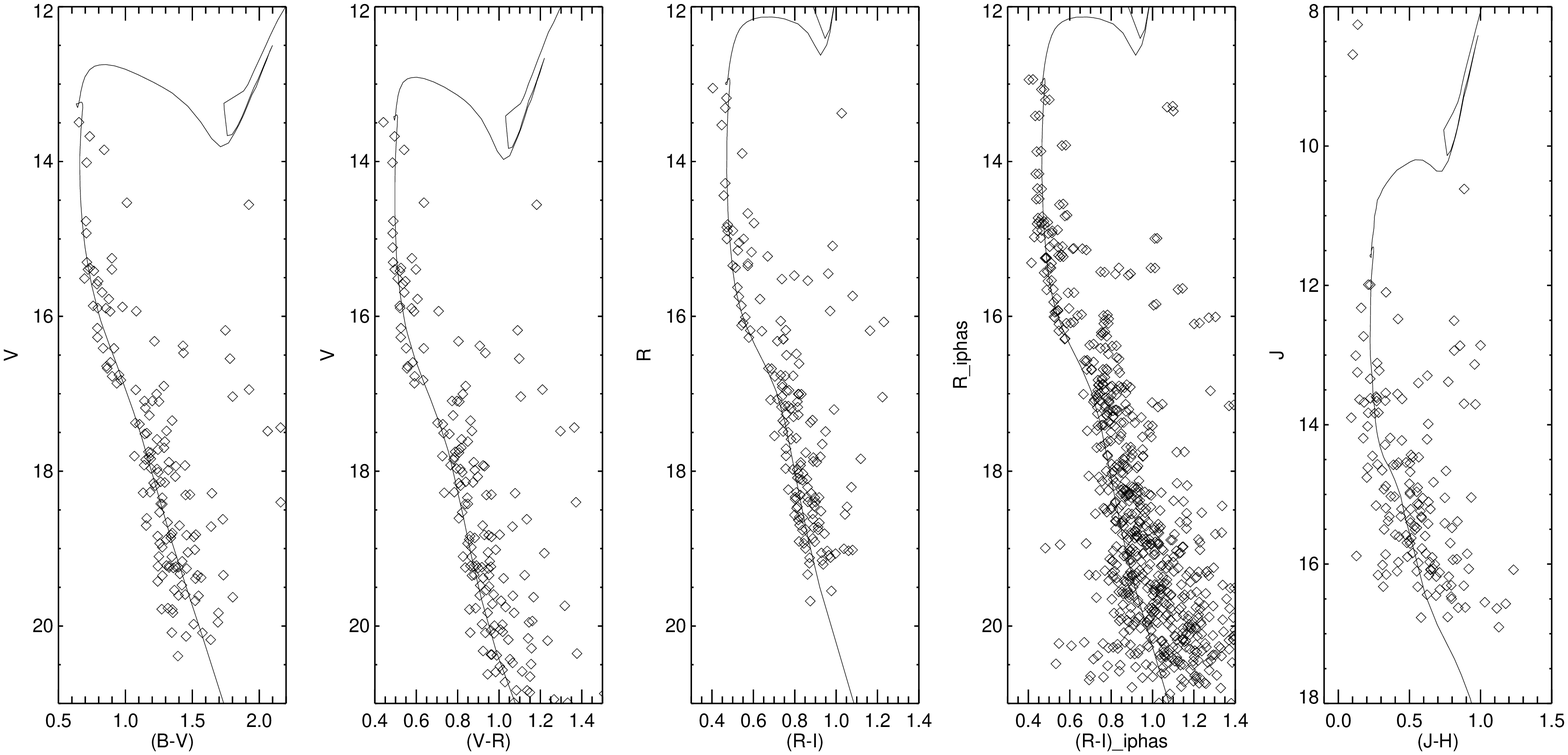}} \caption{
Colour--magnitude diagrams for Berkeley 97. The diamonds mark the
positions of stars on the diagram; the solid lines indicate the
corresponding isochrones. All values along the horizontal and
vertical axes are given in magnitudes. The size of the region for
which the diagram was constructed is $4 \arcmin$.} \label{izo2}
\end{figure*}

\begin{figure*}
\resizebox{\hsize}{!}{\includegraphics{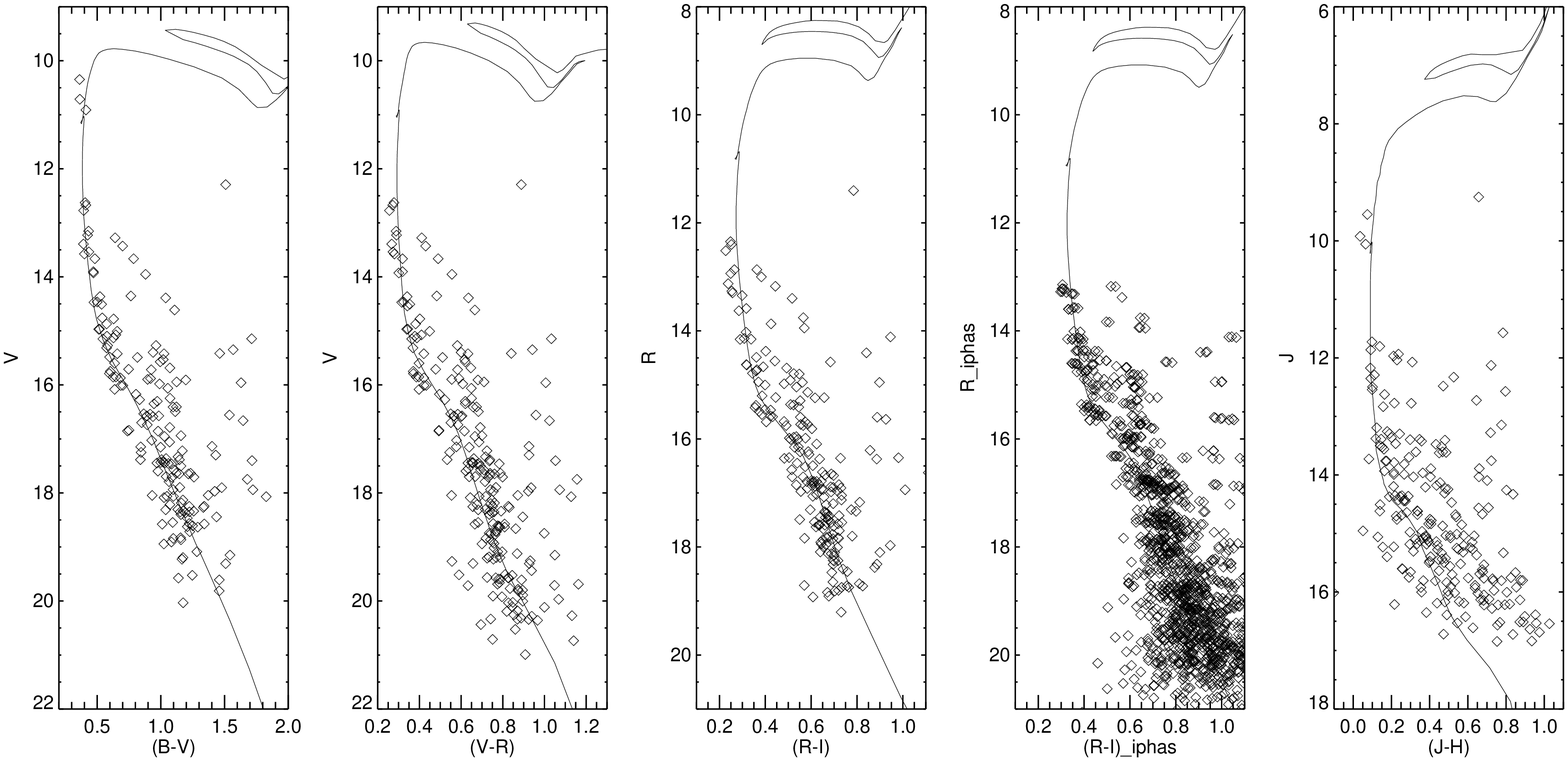}} \caption{
Colour--magnitude diagrams for King 12. The diamonds mark the
positions of stars on the diagram; the solid lines indicate the
corresponding isochrones. All values along the horizontal and
vertical axes are given in magnitudes. The size of the region for
which the diagram was constructed is $5 \arcmin$.} \label{izo3}
\end{figure*}

\begin{figure*}
\resizebox{\hsize}{!}{\includegraphics{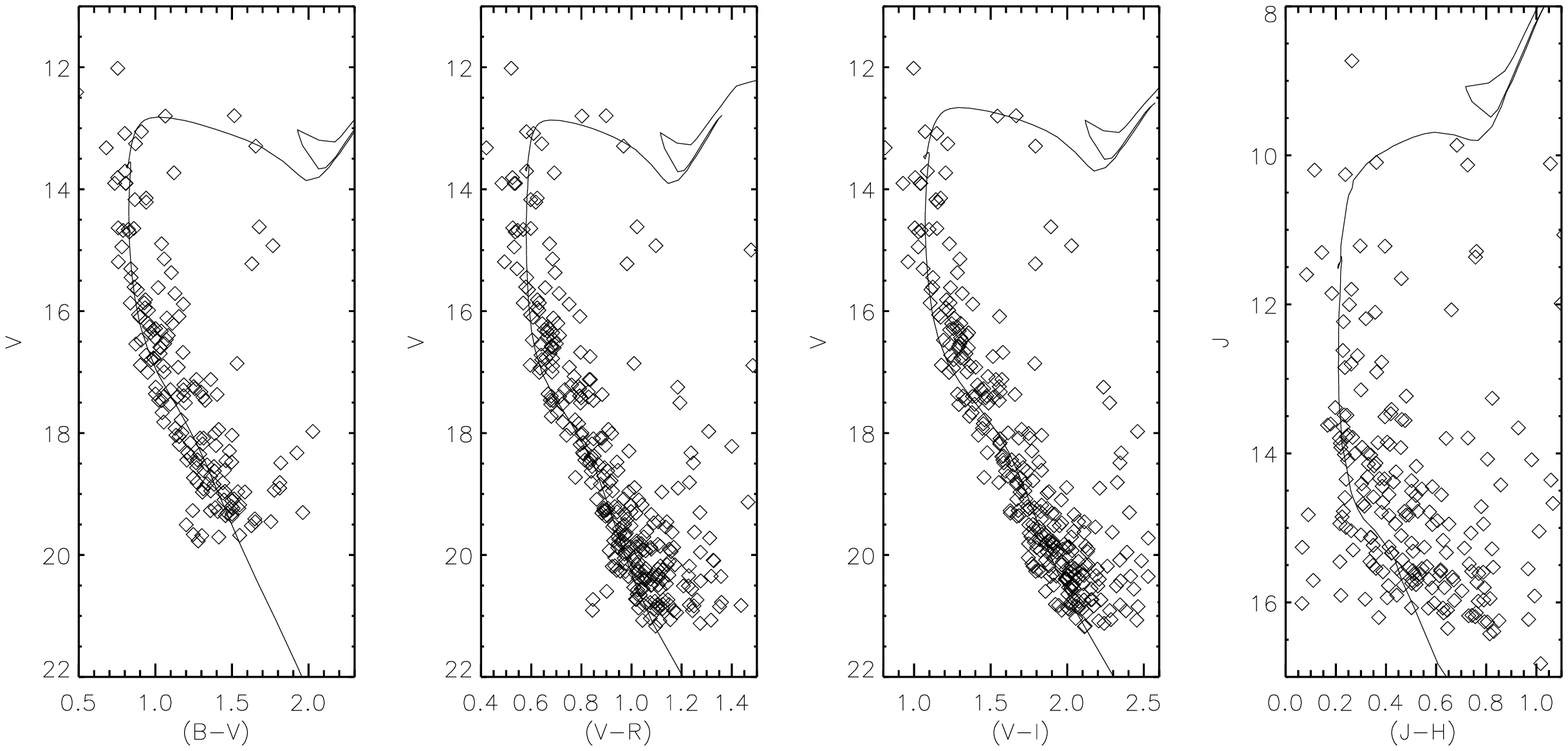}} \caption{
Colour--magnitude diagrams for NGC 7261. The diamonds mark the
positions of stars on the diagram; the solid lines indicate the
corresponding isochrones. All values along the horizontal and
vertical axes are given in magnitudes. The size of the region for
which the diagram was constructed is $7 \arcmin$.} \label{izo4}
\end{figure*}

\begin{figure*}
\resizebox{\hsize}{!}{\includegraphics{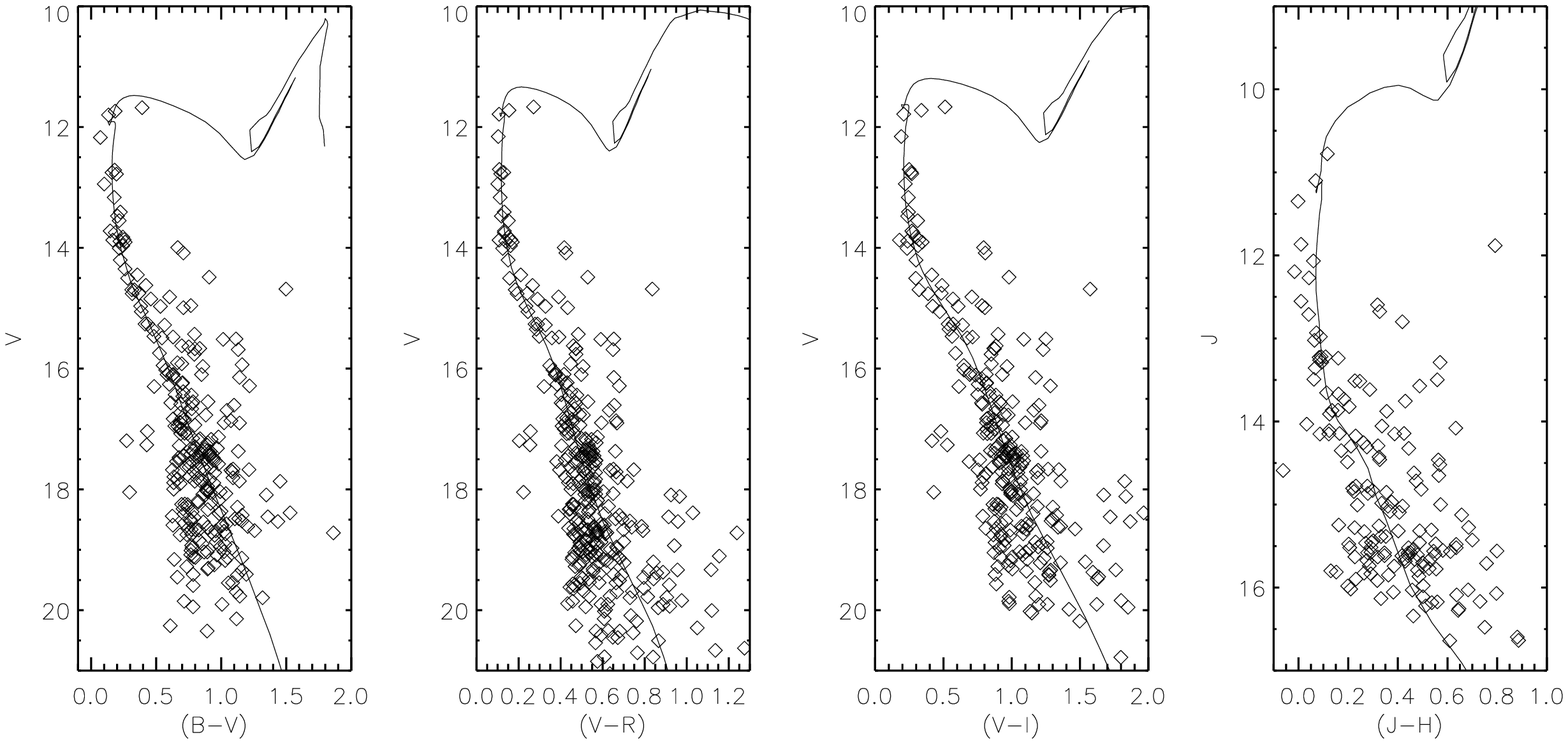}} \caption{
Colour--magnitude diagrams for NGC 7296. The diamonds mark the
positions of stars on the diagram; the solid lines indicate the
corresponding isochrones. All values along the horizontal and
vertical axes are given in magnitudes. The size of the region for
which the diagram was constructed is $8 \arcmin$.} \label{izo5}
\end{figure*}

\begin{figure*}
\resizebox{\hsize}{!}{\includegraphics{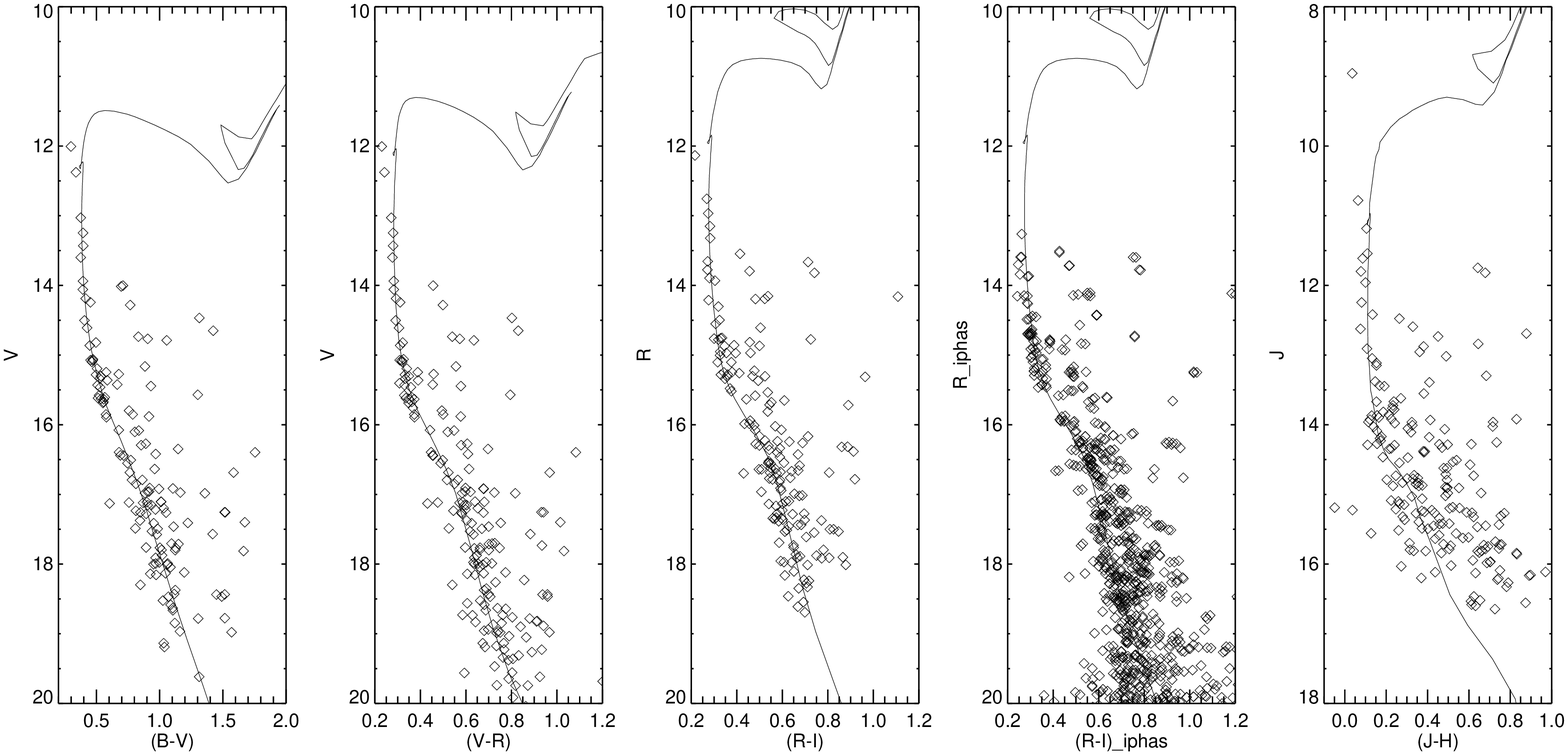}} \caption{
Colour--magnitude diagrams for NGC 7788. The diamonds mark the
positions of stars on the diagram; the solid lines indicate the
corresponding isochrones. All values along the horizontal and
vertical axes are given in magnitudes. The size of the region for
which the diagram was constructed is $4 \arcmin$.} \label{izo6}
\end{figure*}

Figures~\ref{izo1}-\ref{izo6} present the $V -
(B-V)$, $V - (V-R_c)$, $V - (V-I_c)$ (or $R - (R-I_c)$, see explanations below), and $J - (J-H)$
colour--magnitude diagrams, on which the corresponding isochrones are plotted (solid lines).

For each cluster, Table 2 gives the refined coordinates of the
center, $\alpha_{J2000}$ and $\delta_{J2000}$, the cluster
diameter $d$ in arcminutes, the age $t$ in Myr, the colour excesses
$E(B-V)$, and the true distance moduli $(m-M)_0$ determined from each
CMD and recalculated using the relations above. On the whole, the
parameters derived from different colour--magnitude diagrams agree
well, although for some clusters we can see small differences in $E(B-V)$ values estimated separately from optical and infrared data. We suppose that these differences can be due to greater contamination of the infrared CMDs by field stars as compared to optical data, and to the uncertainties of color excess transformations used. The final values of mean colour
excesses and the true distance moduli, along with their $rms$
deviations, are listed in Table 3. All CMDs yield approximately the same age estimates for each cluster, so we did not
put any errors for the ages, which generally depend on the isochrones used. We could not derive the cluster parameters
from the $(K_s,J-K_s)$ diagram for NGC 7788, because of the large dispersion of stars in this CMD.

Let us compare our parameters with those from the literature.
Table 3 contains the colour excesses $E(B-V)$, true distance moduli
$(m-M)_0$, distances $r$ in parsecs, and logarithms of the age
$lgt$ from this paper and different published sources. For the
clusters King 12, NGC 7261, and NGC 7788, we include data from both 
\citet{Loktin2001} and \citet{Dias2002}, because
\citet{Loktin2001} determined the distance moduli, and \citet{Dias2002}
included these distances in his catalogue. Table 3 shows that only
in the case of the King 12 cluster our parameters are in good
agreement with those obtained in the previous study by
\citet{MohanPandey1984}. For the remaining five clusters investigated in this study, the physical parameters have been considerably improved.

\subsection{Mass Function}
For three clusters -- Berkeley 97, King 12, and NGC 7788 -- we
noticed gaps on their main sequences on all CMDs built from our
BV$R_c$$I_c$ CCD photometry. To verify these gaps, we extracted $r$ and $i$ data
from the IPHAS survey (\citet{Drew2005}) and transformed these values
from the SDSS system into Cousins $R$, $I$ magnitudes using
empirical colour transformations by \citet{Jordi2006}. We then built
colour--magnitude diagrams $(R,R-I_c)$ using both our
photometric data and those derived from IPHAS.
Figures ~\ref{izo2},~\ref{izo3}, and~\ref{izo6} include these two CMDs.
For each cluster, the gap on the main sequence is situated at the same
values of magnitudes and colours on both diagrams -- $(R,R-I_c)$ and
$(R_{iphas},(R-I)_{iphas})$. We used solar-metallicity evolutionary tracks and isochrones
by \citet{Girardi2002} to evaluate the mass
intervals (in solar-mass units) that correspond to
these MS gaps: [1.3--1.5] for Berkeley 97, [1.4--1.6] for King 12,
and [1.5--1.7] for NGC 7788.

Present-day luminosity and mass functions for these three clusters
were built by star counts for all stars fainter than those at the MS
turn-off. To evaluate the completeness of our photometric data, we
used the method described by \citet{SagarRichtler1991}.
Artificial stars were added
to our initial images in the $B$ and $I_c$ filters. The fraction of
artificial stars (most of them are faint) does not exceed 10\% of
the total population. Images with artificial stars are then
processed using the same algorithm. The portion of restored stars
in each interval of magnitudes gives us the completeness value in
the same magnitude intervals. We used the least of the two values
obtained by $V$ and $I_c$ images according to \citet{SagarRichtler1991}. This method makes it possible
to find the real PDLF with errors less than 3\% in each magnitude
interval where the completeness is more than 0.5. The members of
each cluster were selected by photometric criteria using $(V,V-I_c)$
diagrams. We built the PDLF for stars inside the
cluster radius and for the field stars, and corrected both functions for the
incompleteness factor and for the area difference. The resulting PDLF
is the difference between the cluster PDLF and the field stars PDLF.

\begin{figure}
 \resizebox{\hsize}{!}{\includegraphics{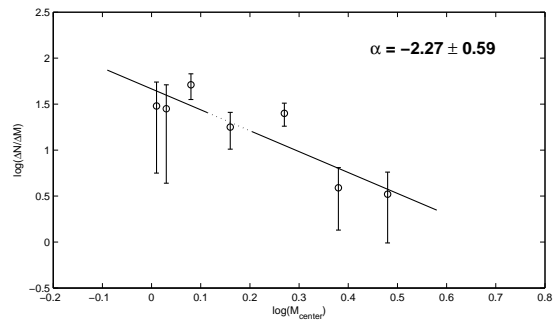}} \caption{
Mass function of Berkeley 97.} \label{izo7}
\end{figure}

\begin{figure}
 \resizebox{\hsize}{!}{\includegraphics{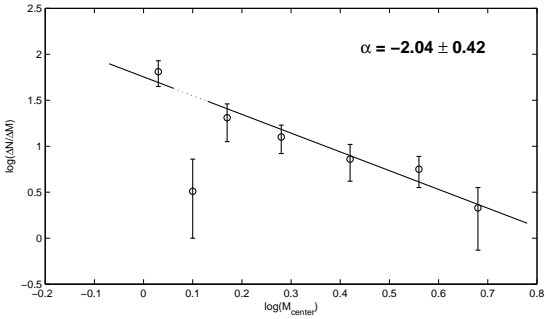}} \caption{
Mass function of King 12.} \label{izo8}
\end{figure}

\begin{figure}
 \resizebox{\hsize}{!}{\includegraphics{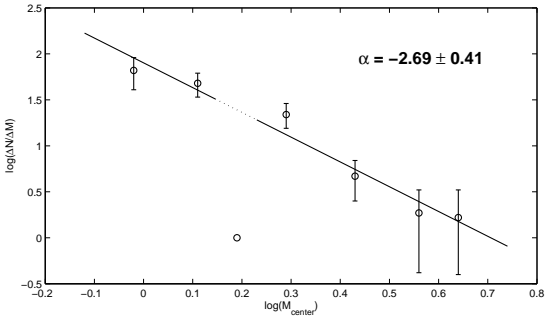}} \caption{
Mass function of NGC 7788.} \label{izo9}
\end{figure}

To convert the luminosity function into the mass function, we used
stellar evolutionary tracks and solar-metallicity isochrones by
\citet{Girardi2002}. Figures~\ref{izo7},~\ref{izo8}, and~\ref{izo9} show
the MFs for clusters Berkeley 97, King 12, and NGC 7788. The values
of the power law exponent (logarithmic mass function slope) --
$\alpha$ was calculated by the $\chi^2$ solution for linear regression
without taking into account escape points in the case of
King 12 and NGC 7788. The gaps on each plot are plotted
according to main sequence gaps on the colour-magnitude diagrams.
The escape point on Fig.~\ref{izo8} and Fig.~\ref{izo9}
corresponds to the deficiency or the absence of stars in the gaps.
There is no such point on Fig.~\ref{izo7} although a main
sequence gap can be clearly seen in the case of Berkeley 97. It can be
explained by the insufficient subtraction of field stars.
Such gaps and breaks in MF plots may be due to a discontinuity in the process
of star formation inside the cluster.

\begin{table*}
\caption{Cluster parameters determined from optical and infrared
data}
\begin{tabular}{lcccccc} \hline
                 & Berkeley 96                  & Berkeley 97                  & King 12                      & NGC 7261                     & NGC 7296                     & NGC 7788\\
\hline
$\alpha_{J2000}$ & $22^h 29^m 49^s$             & $22^h 39^m 28^s$             & $23^h 53^m 01^s$             & $22^h 20^m 07^s$             & $22^h 28^m 01^s$             & $23^h 56^m 38^s$          \\
$\delta_{J2000}$ & $+55\degr23\arcmin47\arcsec$ & $+58\degr59\arcmin51\arcsec$ & $+61\degr56\arcmin45\arcsec$ & $+58\degr07\arcmin41\arcsec$ & $+52\degr19\arcmin22\arcsec$ & $+61\degr24\arcmin02\arcsec$\\
$d$              & $3\arcmin$                   & $4\arcmin$                   & $5\arcmin$                   & $7\arcmin$                   & $8\arcmin$                   & $4\arcmin$                 \\
$t,Myr$              & $40$            & $250$             & $70$                & $160$                  & $280$             & $160$  \\
\multicolumn{7}{c}{$E(B-V)$, mag}\\
$V,B-V$          & $0.50$                       & $0.72$                       & $0.54$                       & $1.00$                       & $0.22$                       & $0.49$                      \\
$V,V-R_c$        & $0.58$                       & $0.85$                       & $0.57$                       & $0.93$                       & $0.24$                       & $0.52$                      \\
$V,V-I_c$        & $0.53$                       & $0.75$                       & $0.51$                       & $0.86$                       & $0.19$                       & $0.49$                      \\
$J,J-H$          & $0.55$                       & $0.79$                       & $0.47$                       & $0.78$                       & $0.28$                       & $0.47$                       \\
$K_s,J-K_s$      & $0.56$                       & $0.72$                       & $0.44$                       & $0.83$                       & $0.25$                       & $--$                       \\
\multicolumn{7}{c}{$(m-M)_0$, mag}\\
$V,B-V$          & $12.71$                      & $12.08$                      & $12.05$                      & $12.16$                      & $12.18$                      & $12.39$                     \\
$V,V-R_c$        & $12.39$                      & $11.86$                      & $11.83$                      & $12.35$                      & $11.97$                      & $12.09$                     \\
$V,V-I_c$        & $12.89$                      & $12.13$                      & $12.19$                      & $12.41$                      & $11.98$                      & $12.30$                     \\
$J,J-H$          & $12.23$                      & $11.76$                      & $11.96$                      & $12.16$                      & $11.72$                      & $12.03$                     \\
$K_s,J-K_s$      & $12.31$                      & $11.71$                      & $11.85$                      & $12.21$                      & $11.92$                      & $--$                     \\
\hline
\end{tabular}
\end{table*}

\begin{table*}
\caption{Comparison of our cluster parameters with published data}
\begin{tabular} {llllll} \hline
            & $E(B-V)$, mag   & $(m-M)_0$, mag   & $r$, pc              & $\lg t$       & Source\\
\hline
Berkeley 96 & $0.54 \pm 0.03$ & $12.51 \pm 0.28$ & $3180^{+440}_{-380}$ & $7.60$        & this paper\\
            & $0.68$          & $13.61$          & $5300$               &               & \citet{Rio1984}\\
Berkeley 97 & $0.77 \pm 0.06$ & $11.91 \pm 0.19$ & $2410^{+220}_{-200}$ & $8.40$        & this paper\\
            & $0.75$          & $11.28$          & $1800 \pm 85$        & $7.30$        & \citet{Tadross2008}\\
King 12     & $0.51 \pm 0.05$ & $11.98 \pm 0.15$ & $2490^{+180}_{-170}$ & $7.85$        & this paper\\
            & $0.52-0.69$     & $11.98$          & $2490 \pm 85$        &               & \citet{MohanPandey1984}\\
            & $0.59$          & $12.034$         &                      & $7.037$       & \citet{Loktin2001}\\
            & $0.59$          &                  & $2378$               & $7.037$       & \citet{Dias2002}\\
            & $0.59$          & $11.88$          & $ $                   & $7.12$        & \citet{Kharchenko2005}\\
NGC 7261    & $0.88 \pm 0.09$ & $12.26 \pm 0.12$ & $2830^{+160}_{-150}$ & $8.20$        & this paper\\
            & $0.48$          & $11.7$           & $2200$               & $8.3$         & \citet{JennensHelfer1975}\\
            & $1.00$          & $12.55$          & $3230$               & $7.0$         & \citet{Fenkart1968}\\
            & $0.969$         & $11.280$         &                      & $7.670$       & \citet{Loktin2001}\\
            & $0.969$         & $ $               & $1681$               & $7.670$       & \citet{Dias2002}\\
NGC 7296    & $0.24 \pm 0.03$ & $11.95 \pm 0.16$ & $2450^{+190}_{-170}$ & $8.45$        & this paper\\
            & $0.15 \pm 0.02$ & $12.33 \pm 0.2$  & $2930 \pm 350$       & $8.0 \pm 0.1$ & \citet{Netopil2005}\\
NGC 7788    & $0.49 \pm 0.02$ & $12.20 \pm 0.17$ & $2750^{+220}_{-210}$ & $8.20$        & this paper\\
            & $0.283$         & $12.030$         &                      & $7.593$       & \citet{Loktin2001}\\
            & $0.283$         & $ $               & $2374$               & $7.593$       & \citet{Dias2002}\\
            & $0.48$          & $11.87$          & $ $                   & $7.48$        & \citet{Kharchenko2005}\\
\hline
\end{tabular}
\end{table*}

\begin{table*}
\caption{Photometric data for Berkeley 96}
\begin{tabular} {llllllllll} \hline
 $\alpha_{J2000}$, deg& $\delta_{J2000}$, deg& $V$  &$\sigma_V$& $B-V$& $\sigma_{(B-V)}$& $V-R_c$& $\sigma_{(V-R_c)}$& $V-I_c$& $\sigma_{(V-I_c)}$   \\
\hline
337.446900 & 55.392340 & 11.200 & 0.003 & 0.326  & 0.005 & 99.999 & 99.999  & 99.999  & 99.999  \\
337.567690 & 55.446010 & 16.511 & 0.027 & 0.806  & 0.014 & 0.525  & 0.026   & 0.990  & 0.024 \\
337.378080 & 55.451410 & 18.611 & 0.028 & 1.145  & 0.062 & 0.722  & 0.036   & 1.314   & 0.040 \\
337.374210 & 55.451490 & 19.292 & 0.023 & 1.394  & 0.036 & 0.926  & 0.016   & 1.720   & 0.049 \\
...        & ...       & ...    & ...   & ...    & ...   & ...    & ...     & ...     & ...    \\ 
\hline
\end{tabular}
\end{table*}

\section{Conclusions}

We used our optical BV$R_c$$I_c$ CCD observations to obtain the $V$
magnitudes and $(B-V)$, $(V-R_c)$, and $(V-I_c)$ colours for a large
number of stars in the fields of six open clusters: Berkeley 96,
Berkeley 97, King 12, NGC 7261, NGC 7296, and NGC 7788 (see
electronic Tables 4-9). For each cluster, we constructed optical
and infrared (based on 2MASS data) colour--magnitude diagrams 
with the Padova isochrones superimposed
(Figs.~\ref{izo1}-\ref{izo6}). We used this technique to derive the ages, distances,
and colour excesses of the clusters under study (Table 3). 

We found three clusters -- Berkeley 97,
King 12, and NGC 7788 -- to have gaps on their main sequences.
Independent data extracted from the IPHAS (\citet{Drew2005}) catalogue confirmed these
features on colour-magnitude diagrams for these three clusters. We
built the luminosity function and the mass function for these clusters and
found the features on the MF plot.

\section*{Acknowledgments}
We used the WEBDA database operated at the Institute for Astronomy
of the University of Vienna and developed by E. Paunzen and J.-C.
Mermilliod; 2MASS data, which is a joint project of the University
of Massachusetts and the Infrared Processing and Analysis
Center/California Institute of Technology, funded by NASA and
NSF; the APASS database, located at the AAVSO website \footnote{http://www.aavso.org/apass}; and the Virtual
observatory resource \footnote{http://vo.astronet.ru} developed at
Sternberg Astronomical Institute (Moscow State University). We thank L.N. Berdnikov, O.V. Vozyakova, and A.K. Dambis for valuable
advice and the anonymous referee for useful comments.



\begin{thebibliography}{}
\bibitem[\protect\citeauthoryear{Becker}{1965}]{Becker1965}
  Becker W.,\ 1965, Mem. Soc. Astron. Italiana, 36, 277
\bibitem[\protect\citeauthoryear{Dias et al.}{2002}]{Dias2002}
 Dias W.S., Alessi B.S., Moitinho A. and Lepine J.R.D.,\ 2002, A\&A, 389, 871
 \footnote{www.astro.iag.usp.br/.wilton/clusters.txt (2012)}
\bibitem[\protect\citeauthoryear{Drew et al.}{2005}]{Drew2005}
 Drew J.E., Greimel R., Irwin M.J., et al.\ 2005, MNRAS, 362, 753
\bibitem[\protect\citeauthoryear{Dutra et al.}{2002}]{Dutra2002}
 Dutra C.M., Santiago B.X. and Bica E.,\ 2002, A\&A, 381, 219
\bibitem[\protect\citeauthoryear{Eggen}{1968}]{Eggen1968}
  Eggen O.J.,\ 1968, Royal Obs. Bull., No. 137
\bibitem[\protect\citeauthoryear{Fenkart}{1968}]{Fenkart1968}
  Fenkart R.P.,\ 1968, Mem. Soc. Astron. Italiana, 39, 85
\bibitem[\protect\citeauthoryear{Frolov}{1977}]{Frolov1977}
  Frolov V.N.,\ 1977, Izv. Glavn. Astron. Obs. Pulkovo, 195, 80
\bibitem[\protect\citeauthoryear{Girardi et al.}{2002}]{Girardi2002}
 Girardi L., Bertelli G., Bressan A., et al.,\ 2002, A\&A, 391, 195
\bibitem[\protect\citeauthoryear{Glushkova et al.}{2010}]{Glushkova2010}
 Glushkova E.V., Zabolotskikh M.V., Koposov S.E. et al., \ 2010, Astronomy
 Letters, 36, 14
\bibitem[\protect\citeauthoryear{Hardie}{1964}]{Hardie1964}
 Hardie R.H., \ 1964, Photoelectric Reductions in the book Astronomical Techniques,
 ed.W. A. Hiltner, University of Chicago, p. 178.
\bibitem[\protect\citeauthoryear{Haug}{1970}]{Haug1970}
  Haug U.,\ 1970, A\&AS, 1, 35
\bibitem[\protect\citeauthoryear{He et al.}{1995}]{He1995}
 He L., Whittet D.C.B., Kilkenny D, and Spencer Jones J.H.,\ 1995,
 A\&AS, 101, 335
\bibitem[\protect\citeauthoryear{Hoag et al.}{1961}]{Hoag1961}
  Hoag A.A., Johnson H.L., Iriarte B., et al.,\ 1961, Publ. US. Nav. Obs., 17,
  345
\bibitem[\protect\citeauthoryear{Jennens \& Helfer}{1975}]{JennensHelfer1975}
  Jennens P.A., Helfer H.L.,\ 1975, MNRAS, 172, 681
\bibitem[\protect\citeauthoryear{Jordi et al.}{2006}]{Jordi2006}
 Jordi K., Grebel E.K. and Ammon A.,\ 2006, A\&A, 460, 339
\bibitem[\protect\citeauthoryear{Kharchenko et al.}{2005}]{Kharchenko2005}
 Kharchenko N.V., Piskunov A.E., Roser S. et al.,\ 2005, A\&A, 438, 1163
\bibitem[\protect\citeauthoryear{Koposov et al.}{2008}]{Koposov2008}
 Koposov S.E., Glushkova E.V. and Zolotukhin I.Yu.,\ 2008, A\&A, 486, 771
\bibitem[\protect\citeauthoryear{Kubinec}{1973}]{Kubinec1973}
 Kubinec W.R.,\ 1973, Publ. Warner Swasey Obs., 1, No. 3
\bibitem[\protect\citeauthoryear{Loktin et al.}{2001}]{Loktin2001}
 Loktin A.V., Gerasimenko T.P. and Malysheva L.K.,\ 2001, Astron.
 Astrophys. Trans., 20, 607
\bibitem[\protect\citeauthoryear{Mohan \& Pandey}{1984}]{MohanPandey1984}
  Mohan V., Pandey A.K.,\ 1984, Ap\&SS, 105, 315
\bibitem[\protect\citeauthoryear{Netopil et al.}{2005}]{Netopil2005}
  Netopil M., Paunzen E., Maitzen H.M., et al.,\ 2005, Astron. Nachr., 326, 734
\bibitem[\protect\citeauthoryear{Del Rio}{1984}]{Rio1984}
  Del Rio G.,\ 1984, A\&AS, 56, 289
\bibitem[\protect\citeauthoryear{Sagar \& Richtler}{1991}]{SagarRichtler1991}
  Sagar R., Richtler T.,\ 1991, A\&A, 250, 324
\bibitem[\protect\citeauthoryear{Skrutskie et al.}{2006}]{2MASS}
  Skrutskie M.F., Cutri R.M., Stiening R., et al.,\ 2006, AJ, 131, 1163
\bibitem[\protect\citeauthoryear{Stetson}{1987}]{Stetson1987}
 Stetson P.B.,\ 1987, PASP, 99, 191
\bibitem[\protect\citeauthoryear{Tadross}{2008}]{Tadross2008}
 Tadross A.L.,\ 2008, MNRAS, 389, 285
\end{thebibliography}
\end{document}